# Lung ultrasound surface wave elastography for assessing instititial lung disease

Xiaoming Zhang, PhD[1], Boran Zhou, PhD[1], Thomas Osborn, MD[2],

Brian Bartholmai, MD[1], Sanjay Kalra, MD[3].

1. Department of Radiology, Mayo Clinic, Rochester, MN 55905, USA
2. Department of Rheumatology, Mayo Clinic, Rochester, MN 55905, USA
3. Department of Pulmonary and Critical Care Medicine, Mayo Clinic, Rochester, MN 55905, USA

*Abstract*
**Lung ultrasound surface wave elastography (LUSWE) is a novel noninvasive technique for measuring superficial lung tissue stiffness. The purpose of this study was to translate LUSWE for assessing patients with interstitial lung disease (ILD) and various connective diseases including systemic sclerosis (SSc). In this study, LUSWE was used to measure the surface wave speed of lung at 100 Hz, 150 Hz and 200 Hz through six intercostal lung spaces for 91 patients with ILD and 30 healthy control subjects. In addition, skin viscoelasticity was measured at both forearms and upper arms for patients and controls. The surface wave speeds of patients' lungs were significantly higher than those of control subjects for the six intercostal spaces and the three excitation frequencies. Patient skin elasticity and viscosity were significantly higher than those of control subjects for the four locations on the arm. In dividing ILD patients into two groups, ILD patients with SSc and ILD patients without SSc, significant differences between each patient group with the control group were found for both the lung and skin. No significant differences were found between the two patients group, although there were some differences at a few locations and at 100 Hz. LUSWE may be useful for assessing ILD and SSc and screening early stage patients.**

## I. INTRODUCTION

Ultrasonography is not widely used for clinically assessing lung disease. Lung tissue is normally filled with air, and the difference in acoustic impedance between air and lung parenchyma is large. Most of the energy of the ultrasound wave is reflected from the lung surface. Ultrasonography evaluation of the thorax is therefore limited to evaluating structures outside of the lung such as pleural fluid, thoracic superficial masses, or adenopathy [1]. However, lung ultrasonography is the standard for diagnosing pleural diseases and is very useful in the emergency and critical care settings [2, 3]. Lung ultrasound imaging typically presents artifacts such as A-lines and B-lines and features such as lung sliding and lung point [3]. These artifacts and features may be used to assess various lung disorders including lung fibrosis [4], pneumothorax [5], and lung consolidations [6]. However, analysis of B-line artifacts is qualitative and relies on visual interpretation which subjects to inter-operator variability [7].

We have developed a lung ultrasound surface wave elastography (LUSWE) technique to measure superficial lung tissue stiffness safely and quickly [8-11]. In LUSWE, a 0.1 second harmonic vibration at a given low frequency is generated by the indenter of a handheld vibrator on the chest wall of a subject. The ultrasound probe is positioned about 5 mm away from the indenter in the same intercostal space to measure the generated surface wave propagation on the lung in that intercostal space. The measurement of surface wave speed on the lung is determined from the change in wave phase with distance and independent of the location of wave excitation.

We are evaluating LUSWE for assessing patients with interstitial lung disease (ILD) in a prospective clinical research study. Patients with ILD have fibrotic and stiff lungs leading to symptoms, especially dyspnea, and may eventually lead to respiratory failure [12]. Many ILDs typically are distributed in the peripheral, subpleural regions of the lung [13, 14]. The superficial distribution of lung fibrosis is especially suited for LUSWE. Diagnosis of lung fibrosis can be difficult, especially early in the disease course, because the symptoms are nonspecific (most commonly shortness of breath and a dry cough) [2, 4, 15]. High-resolution computed tomography (HRCT) is the clinical standard for diagnosing lung fibrosis [16, 17], but it substantially increases radiation exposure for patients. Various HRCT scanning techniques were proposed to reduce the dose [18]. Lung fibrosis results in stiffened lung tissue. However, HRCT does not directly measure lung stiffness.

Both ILD and systemic sclerosis (SSc) are systemic diseases. SSc is a multi-organ connective tissue disease characterized by immune dysregulation and organ fibrosis [19]. Skin involvement in SSc, the major clinical feature, can span from edematous swelling and induration to extensive fibrosis and eventually atrophy. SSc is categorized by variable extent and severity of skin thickening and hardening. The degree of skin involvement is an important measure and predictor of mortality [20]. Severe organ involvement, especially of the skin and lungs, is the cause of morbidity and mortality in SSc [21]. Improvement in skin stiffness is associated with improved survival in many clinical trials [22].

There is a close relationship between lung and skin involvements in both diseases. Patients with SSc have a high incidence of lung fibrosis [19, 23]. Skin and lung fibrosis are prominent features of SSc [19, 24]. A large study showed that 35% of patients with SSc had lung fibrosis and 15% had pulmonary hypertension [23]. Survival rates of patients with SSc-ILD, connective tissue disease–associated ILD, and idiopathic pulmonary fibrosis were studied [25]. Patients with SSc-ILD had better survival rates than those with other ILDs, which was potentially at least partly attributable to routine screening and early detection of ILD in patients with SSc. A multidisciplinary assessment of patients with SSc-ILD may lead to improved diagnosis and management [26, 27].

The Modified Rodnan Skin Score (MRSS) is the standard skin assessment tool in the majority of clinical studies of SSc [28]. It shows that patients with improved MRSS after two years of treatment had improved survival [22]. The MRSS is commonly used as an outcome measure in clinical trials [29]. However, the MRSS is a palpation method, which is subjective, and thus, its accuracy is user-dependent [30]. Moreover, it is difficult to measure the change of skin stiffness over time using palpation [31].

Lung involvement in patients with SSc may be predicted at early stages by evaluation of skin fibrosis because skin manifestation is an early and easily detectable marker of SSc disease activity. Recommendations for future clinical trials stress the critical need for noninvasive, clinically applicable biomarkers to improve the evaluation of patients with SSc-ILD [32]. In this paper, we report our results on both lung and skin stiffness measurements using LUSWE on 91 patients with ILD and 30 healthy control subjects.

## METHOD

### Lung ultrasound surface wave elastography (LUSWE) technique

In LUSWE, a 0.1s harmonic vibration at a frequency is generated on the skin of the chest wall in an intercostal space. The resulting wave propagation at that frequency travels through the intercostal muscle and propagates on the surface of the lung. The wave motions on the selected locations on the lung surface are noninvasively measured using our ultrasound-based method [33]. The phase change with distance of the harmonic wave propagation on the lung surface is analyzed, and from which the surface wave speed is measured,

$$c_s = 2\pi f \left| \Delta r / \Delta \phi \right|, \tag{1}$$

where $\Delta r$ is the distance of two measuring locations, $\Delta \phi$ is the wave phase change over distance, and $f$ is the frequency.

The measurement of wave speed can be improved by using multiple phase change measurements over distances. The regression of the phase change $\Delta \phi$ with distance $\Delta r$ can be obtained by "best fitting" a linear relationship between them, and the equation is

$$\overline{\Delta \phi} = \alpha \Delta r + \beta, \tag{2}$$

where $\overline{\Delta \phi}$ denotes the value of $\Delta \phi$ on the regression for a given distance of $\Delta r$, and α is the regression parameter.

The surface wave speed can be estimated by

$$c_{sr} = 2\pi f \left| \Delta r / \overline{\Delta \phi} \right| = 2\pi f / \alpha, \tag{3}$$

where $c_{sr}$ is the estimation of wave speed from the regression analysis.

The surface wave speed can be related to the elastic modulus of tissue as [34],

$$c_s = \frac{1}{1.05} \sqrt{\frac{\mu}{\rho}} \tag{4}$$

where $\mu$ is the shear elasticity in Pascals and $\rho$ is the mass density of the tissue in kg/m$^3$.

For soft tissue under low frequency harmonic excitation, Voigt's model, which consists of a spring of elasticity $\mu_1$ and a damper of viscosity $\mu_2$ connected in parallel, has been proven to be effective in modeling the linear viscoelastic materials [35-38]. The wave dispersion curve of wave speed $c_s$ with respect to the excitation frequency ω can be formulated by,

$$c_s = \frac{1}{1.05} \sqrt{\frac{2(\mu_1^2 + \omega^2 \mu_2^2)}{\rho(\mu_1 + \sqrt{\mu_1^2 + \omega^2 \mu_2^2})}}. \tag{5}$$

Measurement of lung surface wave speed is noninvasive. Representative LUSWE analysis for a patient is shown in Figure 1(a). The wave motions are measured at eight locations on the lung surface. The normal component of the lung surface motion can be analyzed by cross-correlation analysis of the ultrasound tracking beams [39, 40]. In this study, eight locations over a length of approximately 10 mm on the lung surface were used to measure the normal component of the lung surface motion. The tissue motion is measured at these locations in response to the harmonic wave excitation on the chest wall. A high pulse repetition rate of 2000 pulse/s is used to detect tissue motion in response to the wave excitation at 100, 150, or 200 Hz. A Verasonics ultrasound system (Verasonics, Inc; Kirkland, WA) is used that collects up to a few thousand imaging frames per second by using a plane-wave pulse transmission method.

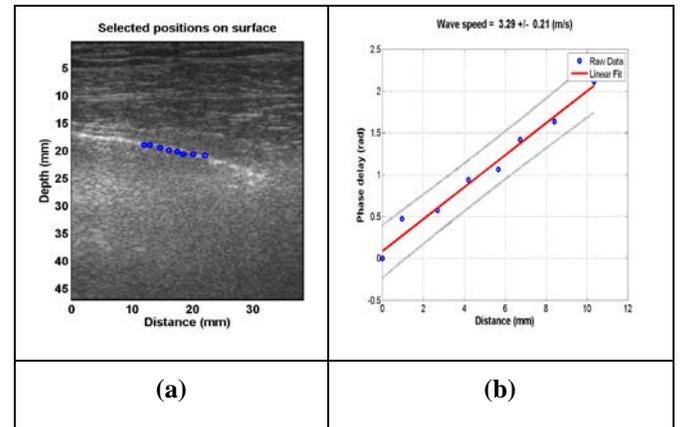

**Figure 1.** (a) Representative LUSWE analysis for a patient subject. The wave motions are measured at eight locations on the lung surface. (b) The wave phase delay of the remaining locations, relative to the first location, is used to measure the surface wave speed.

The surface wave speed on the lung is estimated by determining the change in wave phase with distance along the lung surface. The lung motion at the first location is measured and used as a reference. The wave phase delay of the lung motions at the remaining locations, relative to the reference at the first location, is used to measure lung surface wave speed. The surface wave speed is estimated by the phase changes simultaneously at the 8 locations. Figure 1(b) shows a representative wave speed at 100 Hz for the patient in the left second intercostal space. The surface wave speed was 3.29 ± 0.21m/s (mean ± standard error from the regression analysis) at 100 Hz for the patient. The wave speed on the lung surface is determined by analyzing ultrasound data directly from the lung. Therefore, the wave speed measurement is local and independent of the location and amplitude of excitation.

## Skin measurements

This noninvasive technique can be used to measure viscoelastic properties of other tissues. In this research, we also measured the surface wave speed of skin on the arms at the three excitation frequencies for patients and controls. In skin testing, an ultrasound gel pad standoff by Aquaflex® (Parker Laboratories, Inc., Fairfield, NJ 07004, USA) was placed between the ultrasound probe and the skin to improve the imaging quality of skin. The vibration excitation was directly applied on the skin.

## Patient and healthy control subjects

Human studies were approved by the Mayo Clinic Institutional Review Board (IRB). Each participant completed an informed consent form. Patients were enrolled in this research based on their clinical diagnoses. 91 patients with ILD were enrolled from Mayo Clinic Departments of Rheumatology and Pulmonary and Critical Care Medicine. These patients were confirmed ILD patients with clinical assessments together with pulmonary function test and high resolution CT scans. These ILD patients also had various diseases including systemic sclerosis (SSc) or scleroderma, rheumatoid arthritis, connective tissue disease, idiopathic pulmonary fibrosis, anti-synthetase, Sjögren's syndrome, polymyositis, and systemic lupus erythematosus.

We intended to study the relationship between ILD and SSc by measuring both lung and skin. We divide the patients into two groups, patients with SSc (41 patients) and patients without SSc (50 patients). Patients with SSc were diagnosed according to the American College of Rheumatology (ACR) criteria [41]. The ACR classification criteria for diagnosing SSc requires either 1) the major criterion of proximal scleroderma as judged by palpating or simply observing the skin; or 2) 2 minor criteria such as sclerodactyly, digital pitting scars, or loss of substance from the finger pad, and bibasilar pulmonary fibrosis. Patients' mean age was $62.4 \pm 13.0$ years (range 20-85, 39 male and 52 female). 30 healthy subjects were enrolled as controls if they did not have any lung and skin diseases. Controls' mean age was $45.4 \pm 14.1$ years (range 22-73, 14 male and 16 female).

## Human study protocol

The subject is tested in a sitting position. The subject's lung is tested first and then the skin. It takes about 30-40 minutes to finish the testing. Both lungs of the subject are tested through six intercostal spaces. The upper anterior lungs are tested at the second intercostal space in the mid-clavicular line. The lower lateral lungs are tested at one intercostal space above the level of the diaphragm in the mid-axillary line. The lower posterior lungs are tested at one intercostal space above the level of the diaphragm in the mid-scapular line. Ultrasound imaging is used to identify the lungs and select appropriate intercostal spaces to measure the upper and lower lungs. A 0.1-second harmonic vibration is generated by the indenter of the handheld shaker (Model: FG-142, Labworks Inc., Costa Mesa, CA 92626, USA) on the chest wall. The excitation force from the indenter is much less than 1 Newton and the subject only feels a small vibration on his/her skin. The indenter of the handheld shaker is placed on the chest wall in an intercostal space. The ultrasound probe is positioned about 5 mm away from the indenter in the same intercostal space to measure the resulting surface wave propagation on the lung. An ultrasound probe L11-4 with a central frequency of 6.4 MHz is positioned about 5 mm away from the indenter in the same intercostal space to measure the resulting surface wave propagation on the lung. A Verasonics ultrasound system (Verasonics V1, Verasonics, Inc., Kirkland, WA 98034, USA) is used in this research. Images of the lung and skin are acquired by compounding 11 successive angles at a pulse repetition frequency (PRF) of 2 kHz. The lung is tested at total lung capacity when the subject takes a deep breath and holds for a few seconds. The surface wave speeds are measured at three excitation frequencies of 100 Hz, 150 Hz, or 200 Hz. Three measurements are performed at each location and at each frequency. A small tissue motion in tens of µm is enough for sensitive ultrasound detection of the generated tissue motion. The 100 Hz wave motion is stronger than those of higher frequency waves. The higher frequency waves have smaller wave length but decay more rapidly over distance than the lower frequency waves. The frequency ranges chosen in this study consider the wave motion amplitude, spatial resolution, and wave attenuation. The same ultrasound system and probe are used for skin testing. The subject's arm is placed horizontally on a pillow in a relaxed state. The skin of both left and right forearms and upper arms of subjects are tested. These locations are in the central part of the arms and on the dorsal sides.

## Statistical analysis

An unpaired, two-tailed $t$-test between the patients and healthy control subjects was conducted to compare the sample means. Differences in mean values were considered significant when p<0.05. A one-way ANOVA with critical $p$ value = 0.05 and a *post hoc* Tukey test with $\alpha = 0.05$ was conducted to compare the sample means among three groups: ILD patients with SSc, ILD patients without SSc and healthy control subjects.

## RESULTS

A comparison of wave speeds of lung surface between 91 ILD patients and 30 healthy control subjects is shown in Figure 2 for 100 Hz, 150 Hz, and 200 Hz. The three intercostal spaces are designated by a number from 1 to 3. The upper anterior lung is designated by 1. The lower lungs at the lateral and posterior positions are designated by 2 and 3, respectively. The right and left lungs are designated by letters R and L, respectively. Therefore, L1 represents the left anterior lung in the second intercostal space. The p-values for the $t$-test were less than 0.0001 for all intercostal spaces and for three frequencies between the patients and controls, respectively.

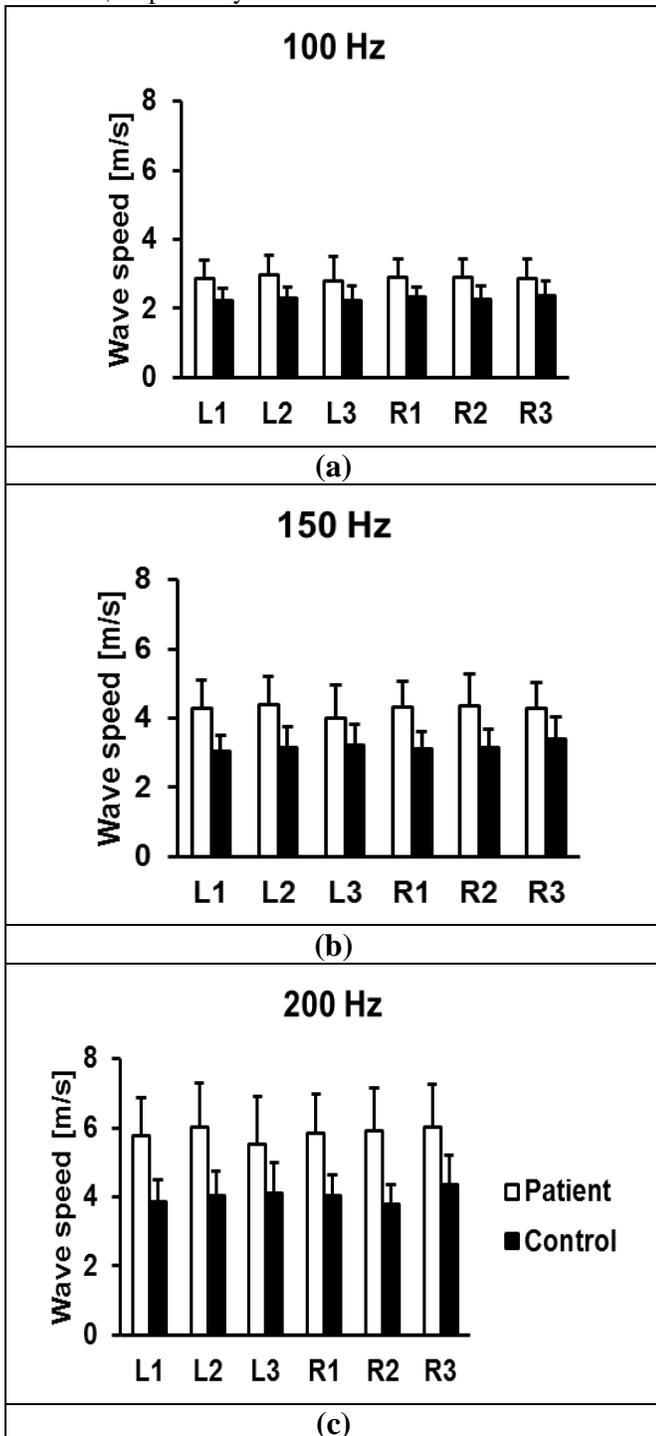

**Figure 2.** Comparison of wave speeds between patients and healthy control subjects through six intercostal spaces. Surface wave speeds were measured at (a) 100 Hz, (b) 150 Hz, and (c) 200 Hz.

Figure 3 shows the comparison of elasticity and viscosity of skin between 91 ILD patients and 30 healthy control subjects. Viscoelasticity is estimated using equation (5) with wave speed measurements at 100 Hz, 150 Hz, and 200 Hz. Most soft tissues have a mass density close to 1.0 g/cm$^3$. In this paper, the mass density of skin is assumed to be 1.0 g/cm$^3$. The forearm and upper arm are designated by numbers 4 and 5, respectively. The right and left arms are designated by letters R and L, respectively. Therefore, R4 represents the skin of the right forearm. The p-values for the $t$-test were less than 0.0001 for the four locations between patients and controls. Therefore, the magnitudes of both elasticity and viscosity of patients were statistically higher than those of healthy subjects.

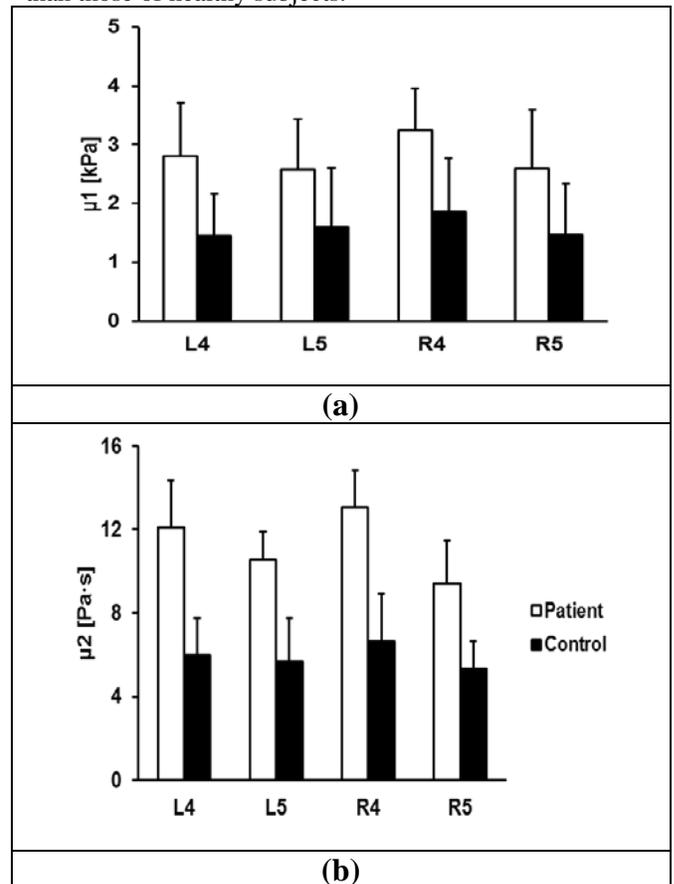

**Figure 3.** Comparison of elasticity μ1 (a) and viscosity μ2 (b) between patients and healthy control subjects at four locations.

One purpose of the research was to study if the relationship between the lung stiffness and the skin stiffness could be identified for patients with ILD or SSc. Both ILD and SSc are systemic diseases. We divide the patients into two groups. Group 1 is the ILD patients with SSc, and group 2 is the ILD patients without SSc. We analyze the date for these two groups of patients and the control subjects.

Figure 4 shows the one-way ANOVA analyses of surface wave speeds of lung for 100 Hz, 150 Hz, and 200

Hz among the three groups: ILD patients with SSc, ILD patients without SSc, and healthy controls.. Table 1 shows the statistical results of lung for the three group subjects at the three frequencies and the six intercostal spaces. The $p$-values for the one-way ANOVA test were less than 0.05 for all intercostal spaces and for three frequencies among the three groups. The *post hoc* Tukey test was performed after ANOVA to analyze the subgroups. Significant differences of wave speed between either group 1 patients or group 2 patients with controls were found. No significant differences were found between the group 1 patients and the group 2 patients for most locations and three frequencies. However, there are statistically significant differences at 100 Hz for locations L1 ($p = 0.003$), R1 ($p = 0.002$) and R3 ($p = 0.001$) between group 1 patients and group 2 patients. There is significant difference at 150 Hz for location R1 ($p = 0.026$) between group 1 patients and group 2 patients. These findings are interesting because ILD affect more on lower lungs but less on upper lungs at L1 or R1. Although we can generally conclude that there are no significant differences of wave speed between ILD patients with SSc and ILD patients without SSc. The upper lungs and low frequencies may provide more information to separate the two group patients.

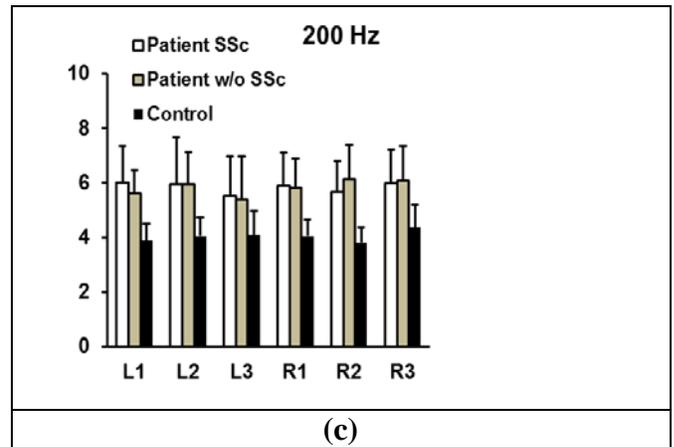

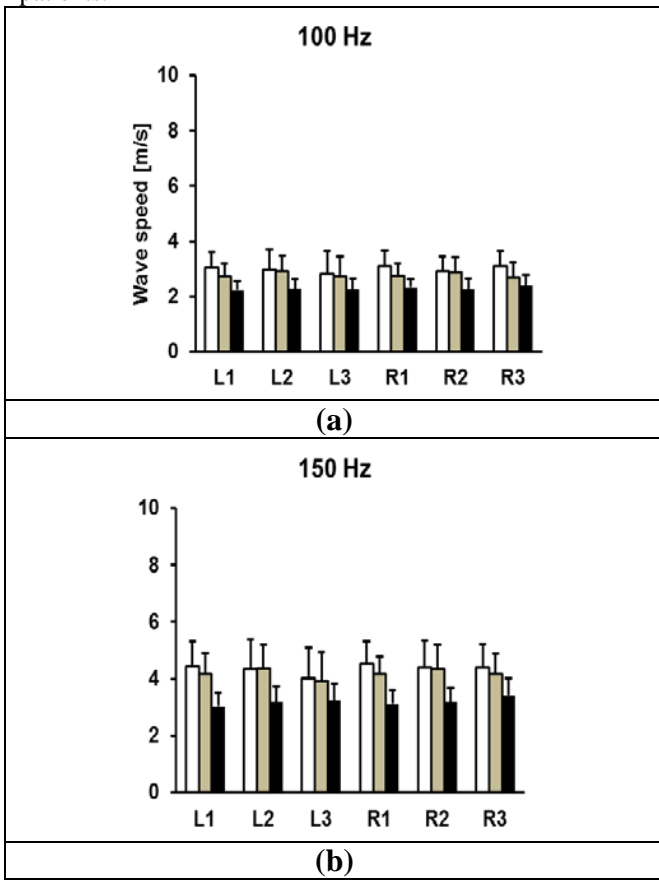

**Figure 4.** Comparison of wave speeds between patients with SSc, patients without SSc, and healthy control subjects through six intercostal spaces. Surface wave speeds were measured at (a) 100 Hz, (b) 150 Hz, and (c) 200 Hz.

**Table 1.** Statistical results of lung surface wave speeds for the three group subjects at the three frequencies and six intercostal spaces. The three groups are patients with SSc, patients without SSc, and healthy control subjects.

| | | | Lung | | | | | |
|---|---|---|---|---|---|---|---|---|
| | | | L1 | L2 | L3 | R1 | R2 | R3 |
| Patient SSc | | mean | 3.05 | 2.98 | 2.81 | 3.09 | 2.90 | 3.09 |
| | | std | 0.57 | 0.72 | 0.85 | 0.59 | 0.55 | 0.56 |
| Patient w/o SSc | 100 Hz | mean | 2.71 | 2.92 | 2.72 | 2.75 | 2.87 | 2.69 |
| | | std | 0.48 | 0.55 | 0.72 | 0.46 | 0.54 | 0.54 |
| Control | | mean | 2.24 | 2.29 | 2.25 | 2.33 | 2.26 | 2.38 |
| | | std | 0.34 | 0.33 | 0.41 | 0.29 | 0.39 | 0.41 |
| | | p | 4.65E-09 | 3.78E-07 | 3.90E-07 | 1.42E-09 | 1.28E-06 | 3.02E-03 |
| Patient SSc | | mean | 4.42 | 4.34 | 4.02 | 4.52 | 4.40 | 4.40 |
| | | std | 0.88 | 1.05 | 1.08 | 0.81 | 0.93 | 0.83 |
| Patient w/o SSc | 150 Hz | mean | 4.17 | 4.36 | 3.93 | 4.18 | 4.35 | 4.18 |
| | | std | 0.74 | 0.82 | 1.01 | 0.60 | 0.86 | 0.68 |
| Control | | mean | 3.03 | 3.16 | 3.23 | 3.12 | 3.16 | 3.41 |
| | | std | 0.48 | 0.58 | 0.60 | 0.49 | 0.51 | 0.62 |
| | | p | 2.92E-14 | 1.08E-09 | 1.66E-07 | 2.88E-12 | 9.70E-09 | 1.61E-03 |
| Patient SSc | | mean | 6.01 | 5.95 | 5.52 | 5.88 | 5.66 | 5.97 |
| | | std | 1.33 | 1.71 | 1.46 | 1.20 | 1.15 | 1.24 |
| Patient w/o SSc | 200 Hz | mean | 5.60 | 5.94 | 5.39 | 5.81 | 6.14 | 6.09 |
| | | std | 0.85 | 1.20 | 1.56 | 1.09 | 1.24 | 1.24 |
| Control | | mean | 3.88 | 4.04 | 4.09 | 4.04 | 3.80 | 4.37 |
| | | std | 0.63 | 0.69 | 0.89 | 0.62 | 0.56 | 0.82 |
| | | p | 2.15E-12 | 1.87E-15 | 2.25E-09 | 7.12E-15 | 1.88E-09 | 4.70E-05 |

Figure 5 shows the one-way ANOVA analyses of elasticity and viscosity of skin among the three groups. Table 2 shows the statistical results of skin for the three group subjects at the four locations on the arm. The $p$-values for the one-way ANOVA test were less than 0.05 for all locations among the three groups. The *post hoc* Tukey tests indicated that significant differences of elasticity or viscosity between either group 1 patients or group 2 patients with controls were found. In addition, no significant differences were found between the group 1 patients and the group 2 patients for most locations. There is one significant difference in viscosity at R4 ($p = 0.019$) between group 1 patients and group 2 patients. We may generally conclude that there are no significant differences of skin elasticity and viscosity between ILD patients with SSc and ILD patients without SSc. Some locations such as R4 may be useful to separate the two group patients.

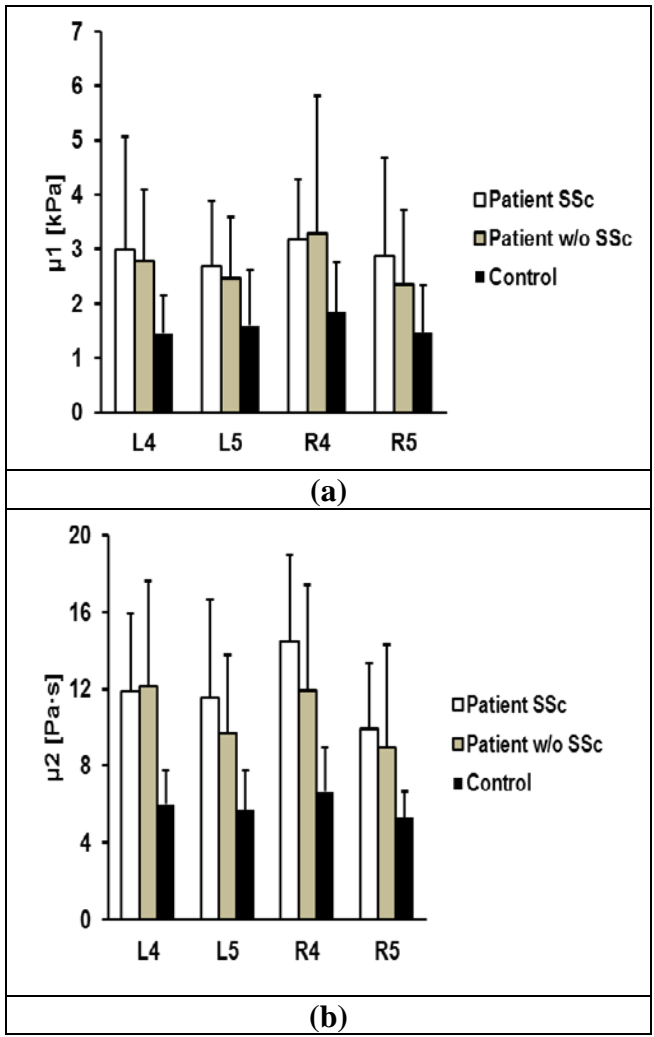

**Figure 5.** Comparison of elasticity µ1 (a) and viscosity µ2 (b) between patients with SSc, patients without SSc, and healthy control subjects at four locations.

**Table 2.** Statistical results of skin elasticity µ1 and viscosity µ2 for the three group subjects at the four locations on the arm.

| | | µ1 [kPa] | | | |
|---|---|---|---|---|---|
| | | Skin | | | |
| | | L4 | L5 | R4 | R5 |
| Patient SSc | mean | 2.98 | 2.69 | 3.17 | 2.87 |
| | std | 2.09 | 1.20 | 1.10 | 1.80 |
| Patient w/o SSc | mean | 2.78 | 2.47 | 3.30 | 2.36 |
| | std | 1.31 | 1.11 | 2.53 | 1.37 |
| Control | mean | 1.45 | 1.59 | 1.85 | 1.47 |
| | std | 0.71 | 1.01 | 0.91 | 0.86 |
| | p | 9.36E-05 | 2.14E-04 | 1.71E-03 | 4.44E-04 |
| | | µ2 [Pas] | | | |
| | | Skin | | | |
| | | L4 | L5 | R4 | R5 |
| Patient SSc | mean | 11.84 | 11.52 | 14.44 | 9.93 |
| | std | 4.11 | 5.12 | 4.51 | 3.39 |
| Patient w/o SSc | mean | 12.14 | 9.72 | 11.91 | 8.94 |
| | std | 5.48 | 4.04 | 5.47 | 5.35 |
| Control | mean | 5.99 | 5.69 | 6.64 | 5.31 |
| | std | 1.76 | 2.05 | 2.27 | 1.32 |
| | p | 9.91E-09 | 1.47E-07 | 4.37E-10 | 1.46E-05 |

## DISCUSSION

LUSWE is a safe technique. Shear wave elastography (SWE) techniques use ultrasound radiation force (URF) to generate tissue motion. However, URF may not be applied to some tissues such as the lung because the relatively high-intensity ultrasound energy may cause alveolar hemorrhage or lung injury [42]. In addition, long period of ultrasound pulses with relatively high-intensity ultrasound energy may cause damage to the ultrasound system itself, e.g., high voltage drop and probe element damage. In LUSWE, the wave generation is safely produced using a gentle mechanical vibration on the skin. URF is not used for generating the shear or surface waves. Diagnostic ultrasound is only used for detection of wave propagation. Therefore, LUSWE can be used for the lung [8, 43] and eye [44, 45].

LUSWE is a simple technique for measuring the skin. Because URF cannot be generated on surface tissue, a standoff ultrasound gel pad is needed for the ultrasound probe to generate and measure the waves on the skin. However, the standoff pad also decays the URF and complicates the URF on the skin. In LUSWE, the mechanical excitation is directly applied on the skin and the standoff pad is only used to improve imaging of the skin.

Optical coherence tomography (OCT) [46] and OCT-based elastography techniques [47, 48] provide high-spatial resolution of skin but cannot measure deep subcutaneous tissue. The imaging penetration of OCT in skin can be up to 1.5 mm [49]. Fibrosis not only affects skin but also subcutaneous tissue [50]. LUSWE can measure much deep subcutaneous tissue than OCT. In the current setup, subcutaneous tissues can be measured to up to 45 mm. LUSWE can be used to assess skin, subcutaneous connective tissue and muscle in the same test which may be useful for assessing these systemic diseases.

Diagnosis of ILD can be difficult at early stages because a patient's symptoms such as shortness of breath and dry cough are nonspecific. The findings of physical examinations are usually nonspecific and chest radiography typically shows nonspecific or nondiagnostic findings. HRCT is the clinical standard for diagnosing lung fibrosis [16, 17], but it is expensive and involves radiation.

Because ILD is a systemic disease, patients enrolled in this research presented with other various diseases. We are most interested in ILD patients with SSc. SSc is a multi-organ connective tissue disease characterized by immune dysregulation and organ fibrosis [19]. Severe organ involvement, especially of the skin and lung, is the cause of morbidity and mortality in SSc. We divided the ILD patients into two groups: ILD patients with SSc (41 patients) and ILD patients without SSc (50 patients). Our results demonstrate that there are significant differences between either patient group with the control group for both lung and skin stiffness. There are not significant differences between ILD patients with SSc and ILD patients without SSc. However, there are some differences between the two patient groups at the upper lungs and at 100 Hz. We plan to study some early stage ILD patients or SSc patients without ILD symptoms. LUSWE may be useful to screen patients at early stages. Connective tissue disease has a high prevalence of ILD, from 7% to 50% for rheumatoid arthritis, Sjogren's syndrome, SSc, and inflammatory myopathies [51, 52]. Testing both lung and skin may provide more information to identify ILD at earlier stage. It takes about 30-40 minutes to finish both lung and skin tests in this research. We may significantly reduce the testing time if we could identify sensitive intercostal spaces and frequencies for assessing specific ILD. However, the current testing time is still good for a clinical study. LUSWE may be easily integrated into a clinical test. LUSWE can be used to quantitatively monitor disease progression for ILD patients. We are continuing to follow the 91 ILD patients, however, a few ILD patients already passed away.

ILD consists of various lung disorders due to damage and fibrosis of the lung parenchyma. HRCT is the clinical standard for diagnosing and characterizing lung fibrosis [16, 17]. Although HRCT provides excellent imaging for assessing lung fibrosis, it does not directly measure lung stiffness. LUSWE provides noninvasive measurement of lung stiffness. LUSWE is a safe, quick and relatively cheap technique for lung testing which may be used for screening patients. In this research, we provide the surface wave speeds at 100 Hz, 150 Hz, and 200 Hz. The elasticity and viscosity of the lung can be estimated from equation (5) if the lung mass density is known. However, we cannot find the data of lung mass density for fibrotic lungs. Also, the lung density is dependent on the pulmonary pressure. In this research, a subject is tested at total lung capacity (TLC) when taking a deep breath and holding. In a paper using an x-ray technique [53], lung density was averaged to be 0.32 $g/cm^3$ for healthy lungs. The measurements were made with the patient sitting either on a chair or in bed and breathing quietly. Lung density was 0.33-0.93 $g/cm^3$ for patients with pulmonary congestion and edema. In an *ex vivo* study on sheep lungs [54], lung density of 0.19-0.26 $g/cm^3$ was used. We expect that the lung density of ILD would be higher than that of healthy lungs. However, there are no data of lung density for ILD patients. Lung mass density is directly associated with lung pathology. Computed tomography (CT) is the major clinical imaging modality for assessing various lung diseases. The mechanism of CT is based on the changes of tissue mass density, but the CT system uses the Hounsfield unit (HU) and not lung density to image the lung. We are developing a deep neural network (DNN) model to predict lung mass density on lung phantoms based on the LUSWE measurements [55]. We will study how to develop and apply the DNN models for the patient's data.

LUSWE is used in this study to evaluate ILD, because ILD is mostly superficial tissue disease. However, LUSWE may be useful for assessing other lung diseases. Pulmonary edema is a fundamental feature of congestive heart failure and inflammatory conditions such as acute respiratory distress syndrome [56]. The presence of extravascular lung water (EVLW) predicts a worse prognosis in critically ill patients [57] and increased risk of death or heart failure readmission [58]. In the presence of EVLW, the ultrasound beam finds subpleural interlobular septa thickened by edema [59]. The reflection of the ultrasound beam produces reverberation artifacts, called "B-lines" in wet lungs. The features of "B-lines" are used to evaluate pulmonary edema [60] and congestive heart failure [61]. A recent multicenter trial demonstrated that lung ultrasound (LUS) was more sensitive than chest x-ray for the diagnosis of cardiogenic pulmonary edema [62]. However, analysis of B-line artifacts is qualitative and relies on visual interpretation which is subject to inter-operator variability [7]. We have started to evaluate LUSWE for assessing EVLW on patients with heart failure.

We have found significant differences for both lung and skin stiffness between the patients and controls. One of the reasons may be due to the fact that the patient's group is relative older than the control's group. We will plan to study the control group with older age around 60.

## CONCLUSION

Lung ultrasound surface wave elastography is a novel noninvasive technique for measuring superficial lung tissue stiffness. In this study, LUSWE was used to measure both lung and skin for 91 patients with ILD and 30 healthy control subjects. The surface wave speeds of patients' lungs were significantly higher than those of control subjects for the six intercostal spaces and three excitation frequencies. Patient skin elasticity and viscosity were significantly higher than those of control subjects for the four locations on the arm. In dividing ILD patients into two groups, ILD patients with SSc and ILD patients without SSc, significant differences between each patient group with the control group were found for both the lung and skin. No significant differences were found between the two patient groups although there were some differences at a few locations and at 100 Hz. LUSWE may be useful for assessing ILD and SSc and screening early stage patients.

### ACKNOWLEDGMENT

This study is supported by NIH R01HL125234 from the National Heart, Lung, and Blood Institute.